\documentclass{camera-hepex}
\usepackage{epsfig}

\newcommand{\ee}{\mbox{$\mathrm{e}^{+}\mathrm{e}^{-}$}}

\def\be{\begin{equation}}
\def\ee{\end{equation}}
\def\bea{\begin{eqnarray}}
\def\eea{\end{eqnarray}}

\begin{document}
\unitlength 1cm \vspace*{1.cm} \title{HINTS OF HIGGS BOSON \\
  PRODUCTION AT LEP}

\author{Marumi Kado}

\organization{European Organization for Nuclear Research (CERN),\\
1211 Geneva 23, Switzerland.}

\maketitle 

\vspace{-1cm}

{\abstract An excess of signal-like events above the expected
  background, corresponding to appro\-ximately three standard
  deviations, was observed in the search for the standard model Higgs
  boson at LEP in 2000. This excess is consistent with the existence
  of a 115~GeV/c$^2$ Higgs particle.} 

\vspace{2cm}

The LEP run in 2000 at centre-of-mass energies up to 209~GeV allowed
the standard model Higgs boson search sensitivity at three standard
deviations ($\sigma$) to extend up to a mass of 115~GeV/c$^2$. Around
this limit of sensitivity, an excess of 2.9$\sigma$ was
observed~\cite{lepc}. Its characteristics are consistent with those
expected from the signal: the distributions of the events among the
experiments, among the decay channels, as a function of time, and as a
function of signal purity match the signal hypothesis~\cite{request}.
A complete introduction on the generalities of Higgs boson searches at
LEP will not be given here and can be found in~\cite{aleph,delphi,l3,opal}. The
results given herein are based on the $\sim$810~pb$^{-1}$ of data used
for the latest LEP wide combination~\cite{lepc}, among which
$\sim$490~pb$^{-1}$ were taken at energies above 206~GeV. This data
sample is lacking $\sim$30~pb$^{-1}$ of data collected in the last
week before the final shutdown of LEP. However all collaborations have
published results on their entire data sample.

\vspace{.25cm}
\noindent{\bf {\it Results of the Individual Experiments}}
\vspace{.25cm}

\begin{figure}[t]
\begin{center}
\begin{tabular}{cc}
\mbox{\epsfxsize=.437681\hsize\epsfysize=.1915\vsize\epsfbox{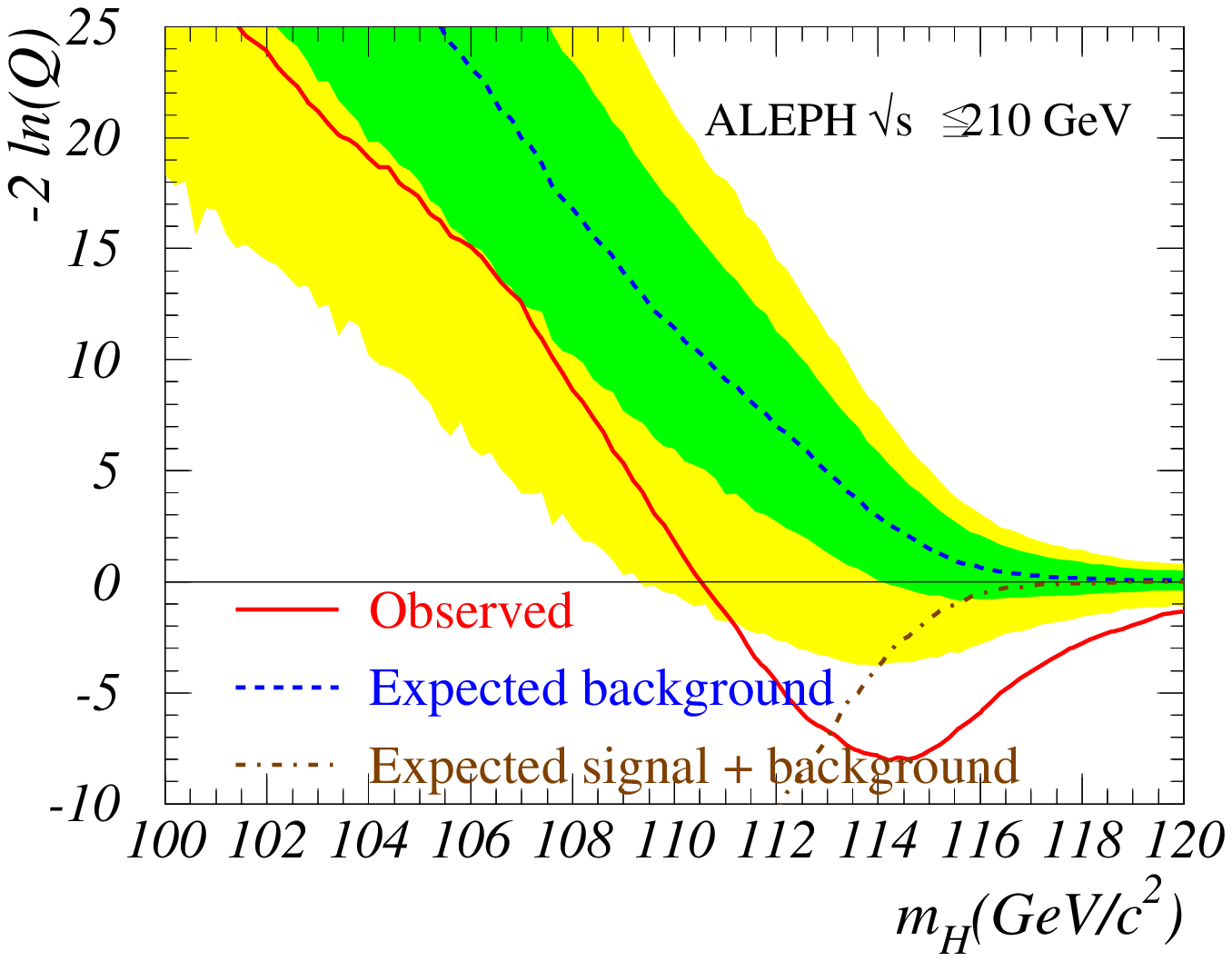}} 
\put(-1.85,3.5){(a)} &
\hskip.45cm\mbox{\epsfxsize=.4\hsize\epsfbox{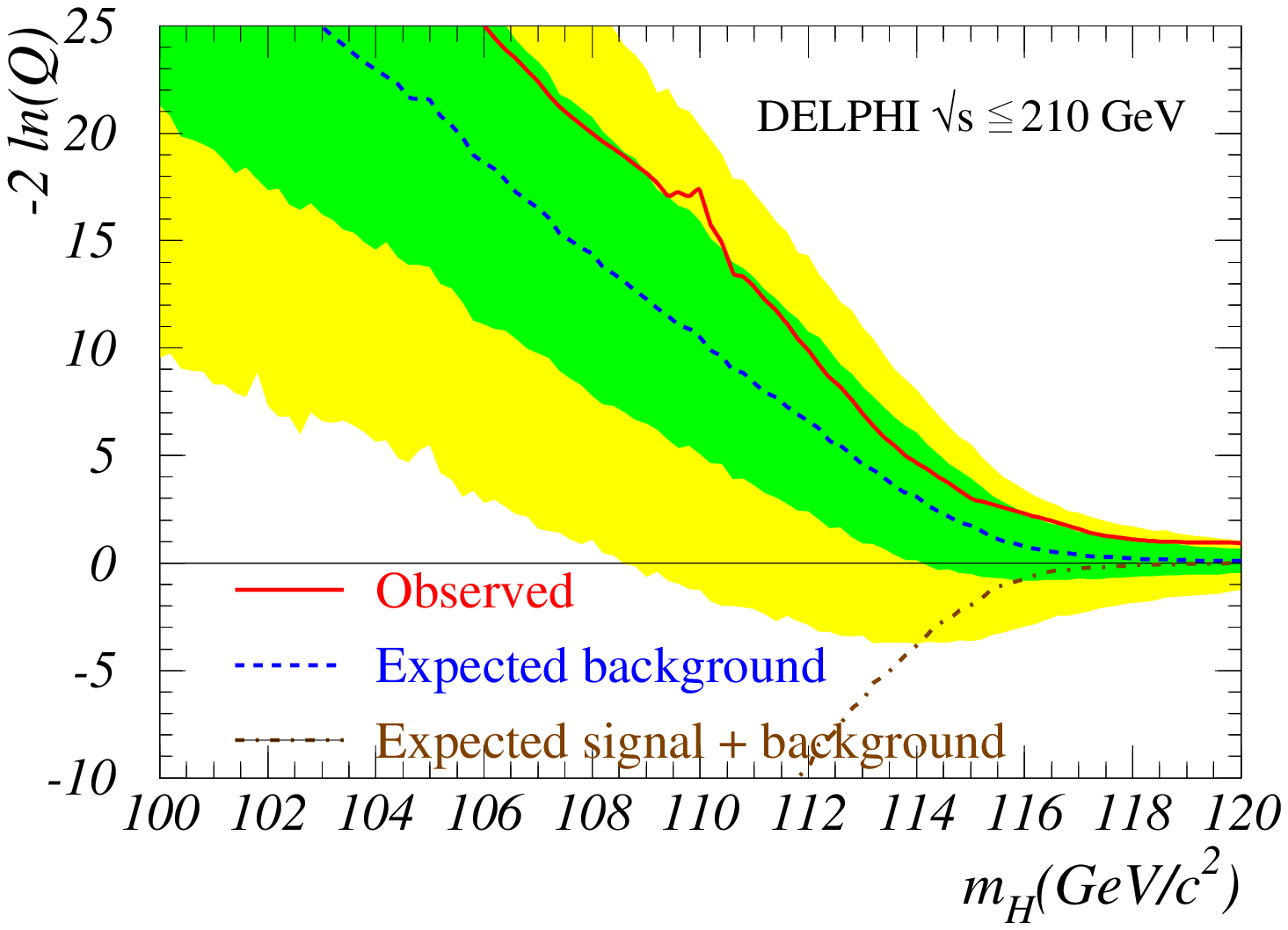}} 
\put(-1.35,3.5){(b)} \\
\hskip-.45cm \mbox{\epsfxsize=.4\hsize\epsfbox{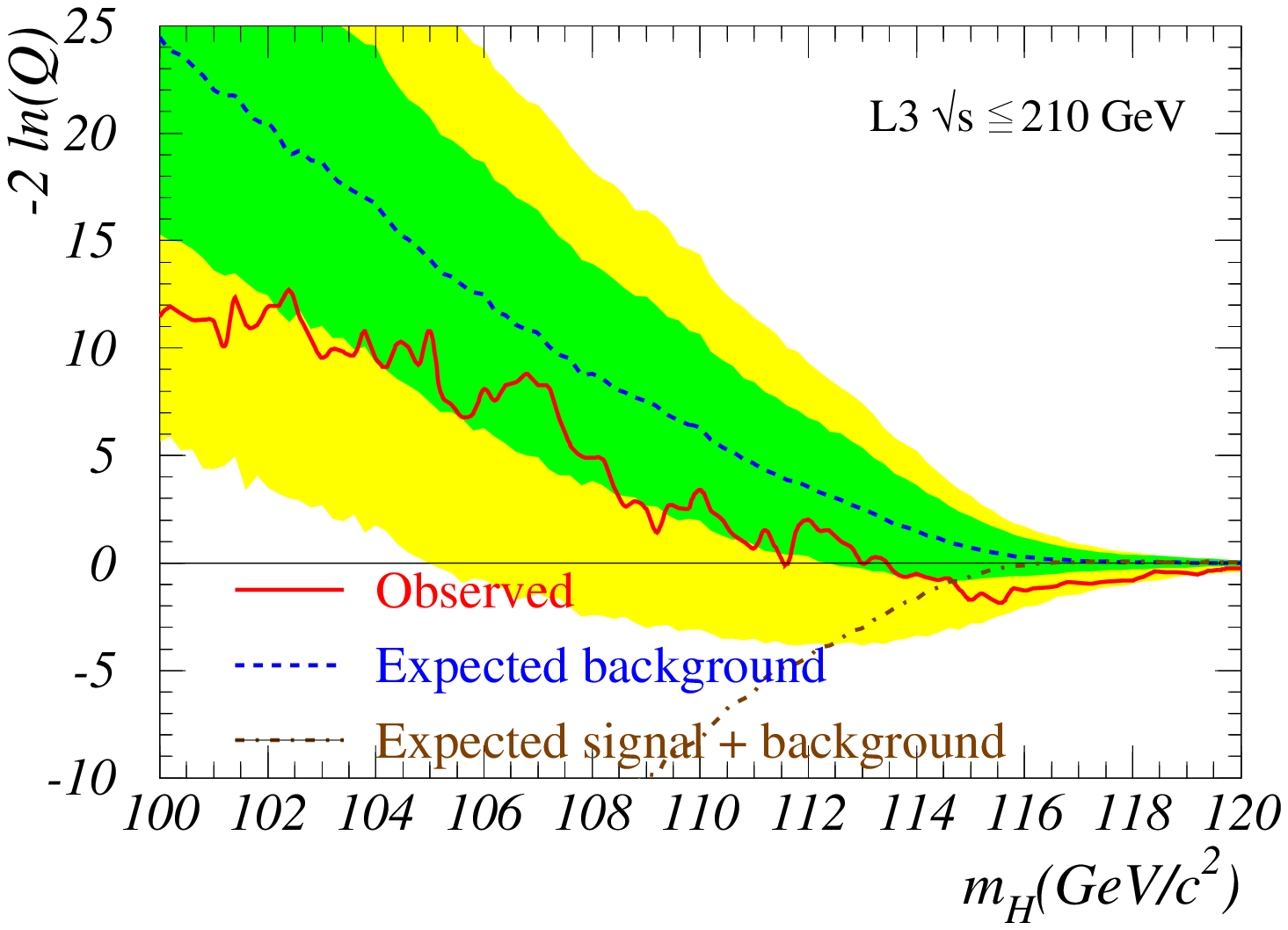}} 
\put(-1.35,3.5){(c)} &
\hskip.45cm \mbox{\epsfxsize=.4\hsize\epsfbox{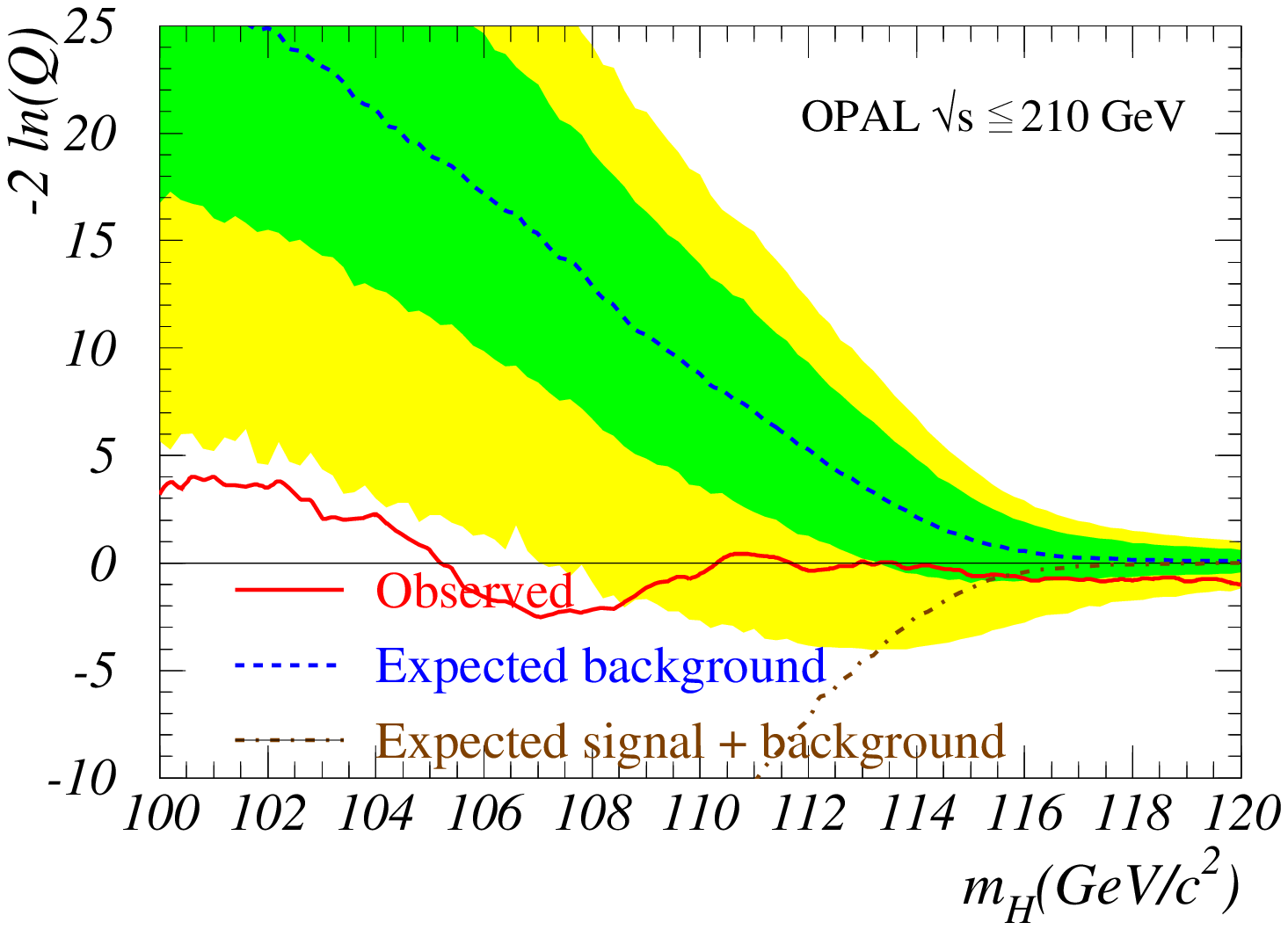}}
\put(-1.35,3.5){(d)}
\end{tabular}
\vspace{-.2cm}
\caption{The log-likelihood estimator for the (a) ALEPH, (b) DELPHI, (c) L3 and (d) OPAL 
  experiments as a function of the Higgs boson mass hypothesis, for
  the observation (solid), background only (dashed) and signal (dashed
  dotted) expectations. The dark and light gray regions around the
  background only expectation illustrate the one and two sigma bands
  respectively.
\label{experiments}}
\end{center}
\end{figure}

To quantify the result of a search channel, the full discriminating
power of certain event characteristics, such as reconstructed mass,
b-tagging or a final discriminating variable ({\it e.g.} obtained with
neural network or likelihood techniques), are all taken into account
in a log-likelihood estimator $-2\ln Q$ of a Higgs boson mass
hypothesis ($m_{\rm H}$). The latter is defined as:
$$-2\ln Q(m_{\rm H}) = 2s_{tot}(m_{\rm H})-2\sum_{i} \ln
(1+\frac{s_i}{b_i}(m_{\rm H}))$$
where $s_{tot}$ is the total amount
of signal expected in the channel and $s_i/b_i$ is the ratio of the
signal and the background values of the distribution of the
discriminant quantities used for the $i^{\rm th}$ candidate event
observed; this ratio will be referred to as event weight. Independent
channels and experiments are simply combined by adding their
estimators. The distribution of the estimator for each individual
experiment is shown in Fig.~\ref{experiments} as a function of the
Higgs boson mass hypothesis. The result obtained by the ALEPH search
shows a clear minimum at high signal mass hypotheses, mostly
due to an excess of four jet events with reconstructed masses near the
kinematic threshold. The most likely signal mass hypothesis is
114~GeV/c$^2$. The probability (1-CL$_{\rm b}$) for this excess to be
due to a fluctuation of the background when testing a mass
hypothesis of 114~GeV/c$^2$ is 0.15\%. The mass hypothesis for which
the excess is most unlikely to result from a background fluctuation is
116~GeV/c$^2$ with 1-CL$_{\rm b}$$\sim$0.05\%.  This hint of the
existence of a Higgs boson with mass around 115~GeV/c$^2$ is further
supported by the observations of L3 and OPAL. Though less
significantly, both experiments observe an excess of events at high
reconstructed masses with respective probabilities of 6.8\% and 19\%.
DELPHI observes a deficit of events which is compatible with the
background only hypothesis for a signal mass hypothesis of
115~GeV/c$^2$ (1-CL$_{\rm b}$=67\%) but not incompatible with the
signal hypothesis.  The probability (CL$_{\rm s}$) that this deficit
is due to a downward fluctuation of a 115~GeV/c$^2$ signal is
$\sim$14\%.  Although most of the effect is seen in ALEPH, the
distribution of the excess among the four experiments is consistent
with the expectation in the presence of a $\sim$115~GeV/c$^2$ Higgs
boson.

\vspace{.25cm}
\noindent{\bf {\it Combined Result}}
\vspace{.25cm}

\begin{figure}[t]
\begin{center}
\begin{tabular}{ccc}
\hskip-.9cm \mbox{\epsfxsize=.38\hsize\epsfbox{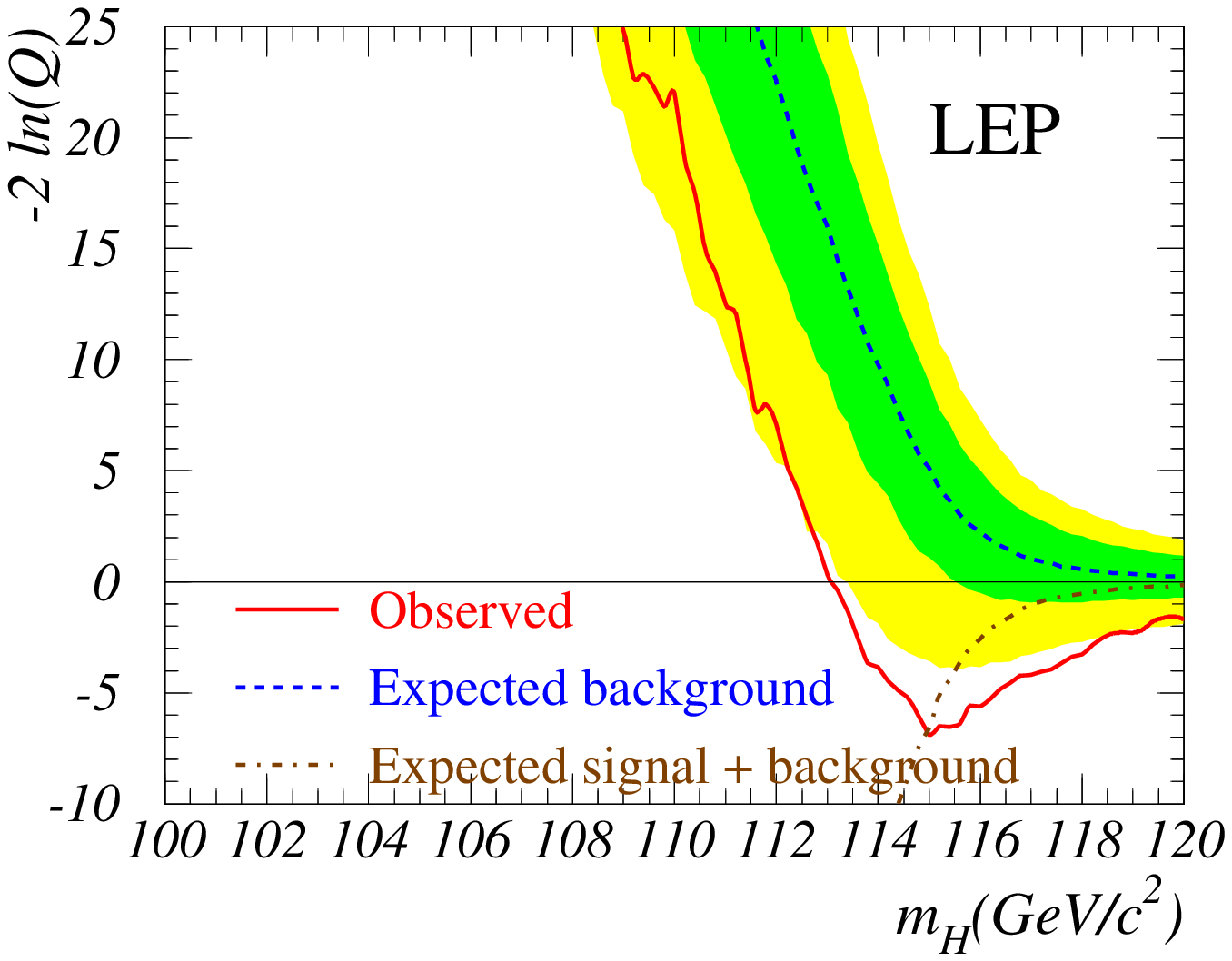}} 
\put(-1.5,3.){(a)} &
\hskip-.75cm \mbox{\epsfxsize=.38\hsize\epsfbox{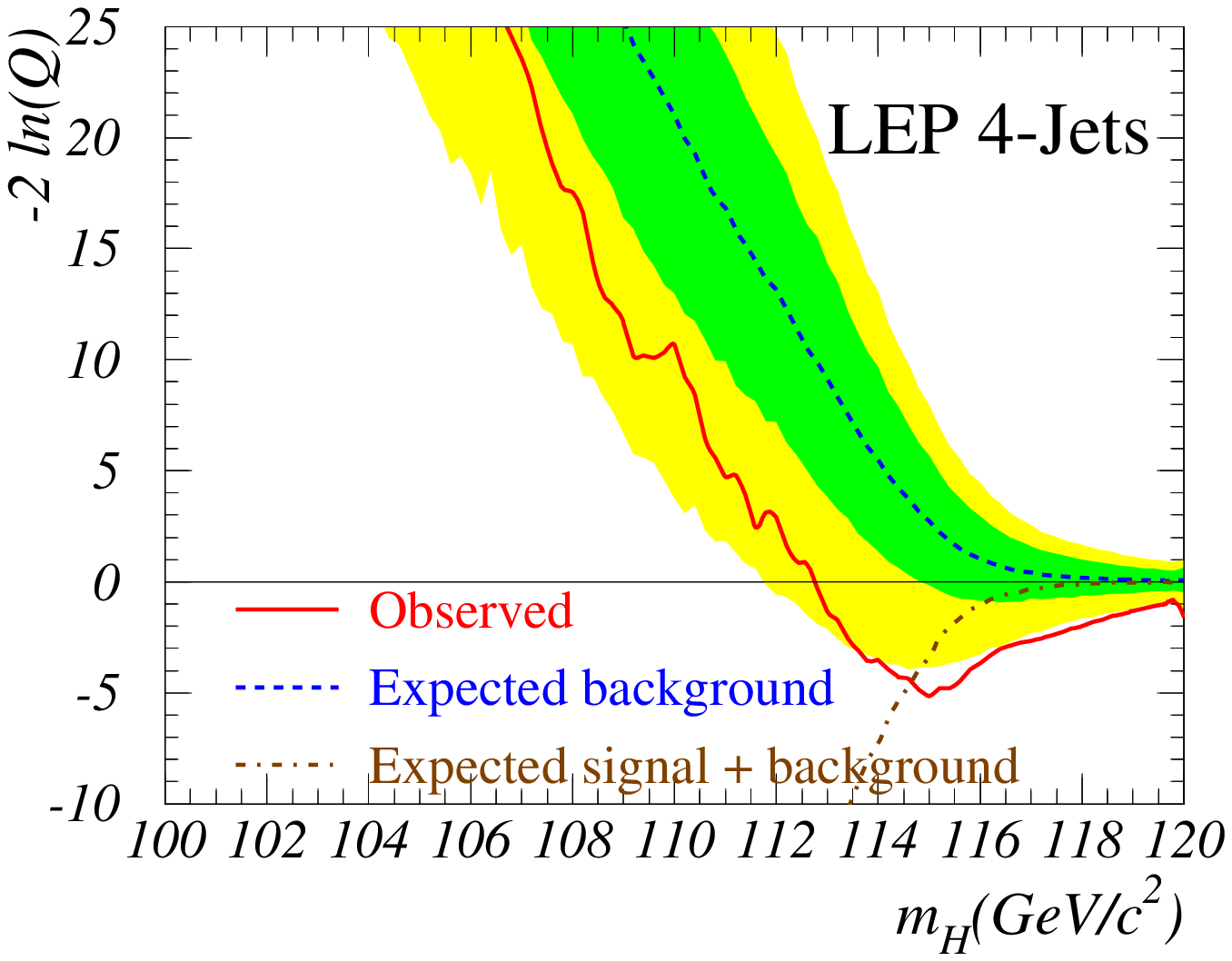}} 
\put(-1.5,3.){(b)}
\hskip-.4cm \mbox{\epsfxsize=.38\hsize\epsfbox{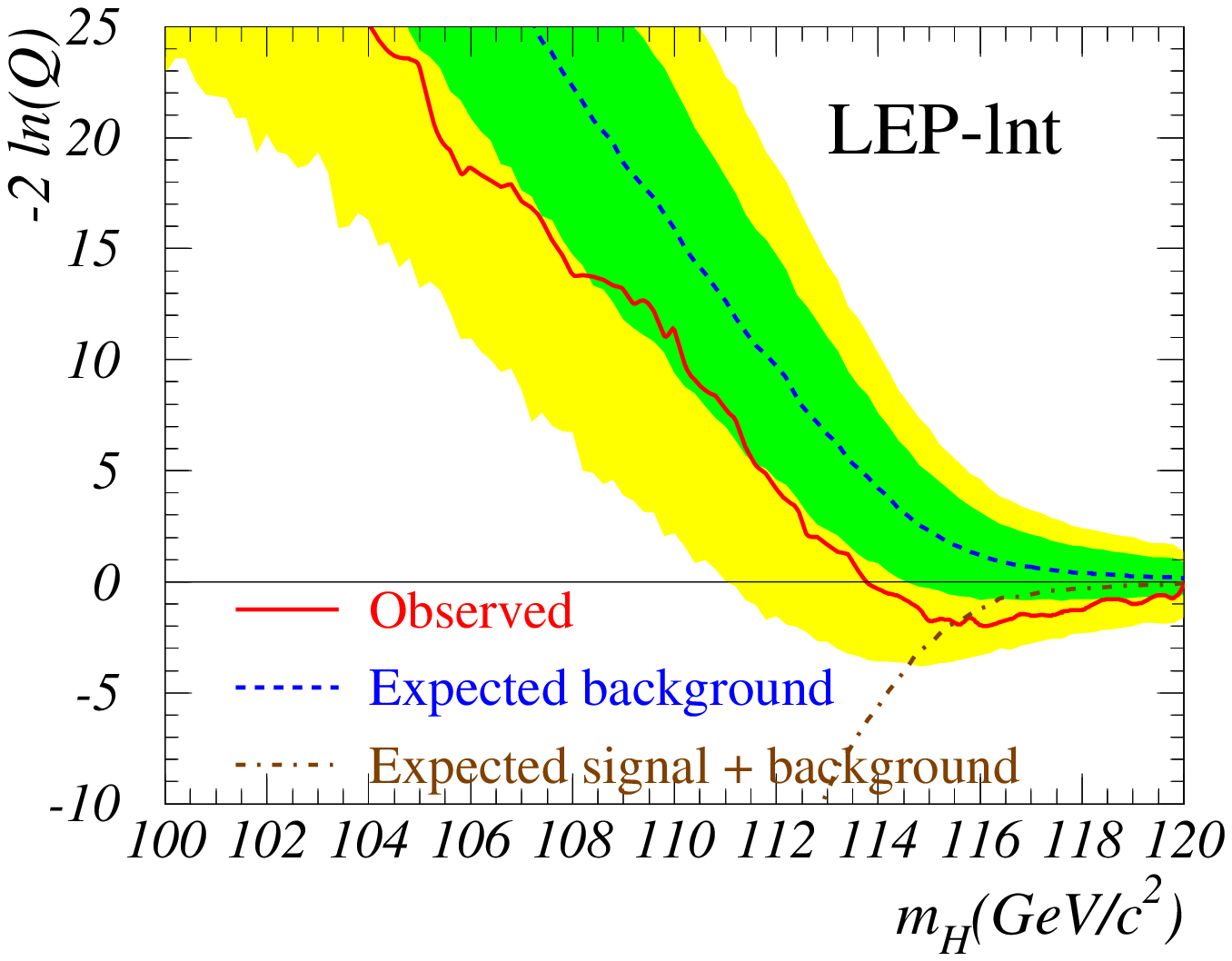}} 
\put(-1.5,3.){(c)}
\end{tabular}
\vspace{-.2cm}
\caption{The log-likelihood estimator for the (a) combination of all 
  channels and experiments, (b) the combination of the four jet
  channels and (c) of all channels except the four jet of all
  experiments as a function of the Higgs boson mass hypothesis, for
  the observation (solid), background only (dashed) and signal (dashed
  dotted) expectations.
\label{combination}}
\end{center}
\end{figure}

This consistency is reflected in the combination of the four
experiments where the most likely Higgs boson mass hypothesis is
115~GeV/c$^2$, in agreement with the signal expectation, as seen in
Fig.~\ref{combination}-a. The combined probability that the excess be
due to a background fluctuation is 0.042\%, corresponding to a
significance of 2.9$\sigma$. Selecting the most signal like events by
requiring that their event weight for a Higgs boson mass hypothesis of
115~GeV/c$^2$ be larger than 0.3, 14 event are observed in the data, 7
events are expected from all background processes and 7 are expected
from a 115~GeV/c$^2$ signal. These candidate events are well
distributed among experiments (6 are selected in ALEPH, 3 in L3, 3 in
OPAL and 2 in DELPHI) and among the search channels (9 are selected in
the four jet channel, 3 in the missing energy channel and two in the
charged lepton channels). These numbers are consistent with the signal
hypothesis.

\vspace{.25cm}
\noindent{\bf {\it Consistency and Robustness}}
\vspace{.25cm}

Many systematic studies were performed to substantiate the
compatibility of the observation with the signal hypothesis. Three
relevant examples are given here.

The four quark final state is considerably more sensitive to the
signal than the leptonic (electron, muon, tau or neutrino) final
states. However, the combination of all leptonic channels performs
nearly as well as the four quark channel. The results of these
combinations are illustrated in Fig.~\ref{combination}-b and c. In
both cases an excess is observed with a minimum of the estimator at
masses around $\sim$115~GeV/c$^2$, showing that the effect is shared
between channels. The probabilities for these excesses to be due to
background fluctuations are, in terms of standard deviations,
2.3$\sigma$ and 1.9$\sigma$ respectively for the combinations of the
four jet channel at 115~GeV/c$^2$ and the leptonic channels at
116~GeV/c$^2$. This check confirms the consistency of the excess amongst
channels.

\begin{figure}[t]
\begin{center}
\begin{tabular}{cc}
\mbox{\hskip.3cm\epsfxsize=.45\hsize\epsfbox{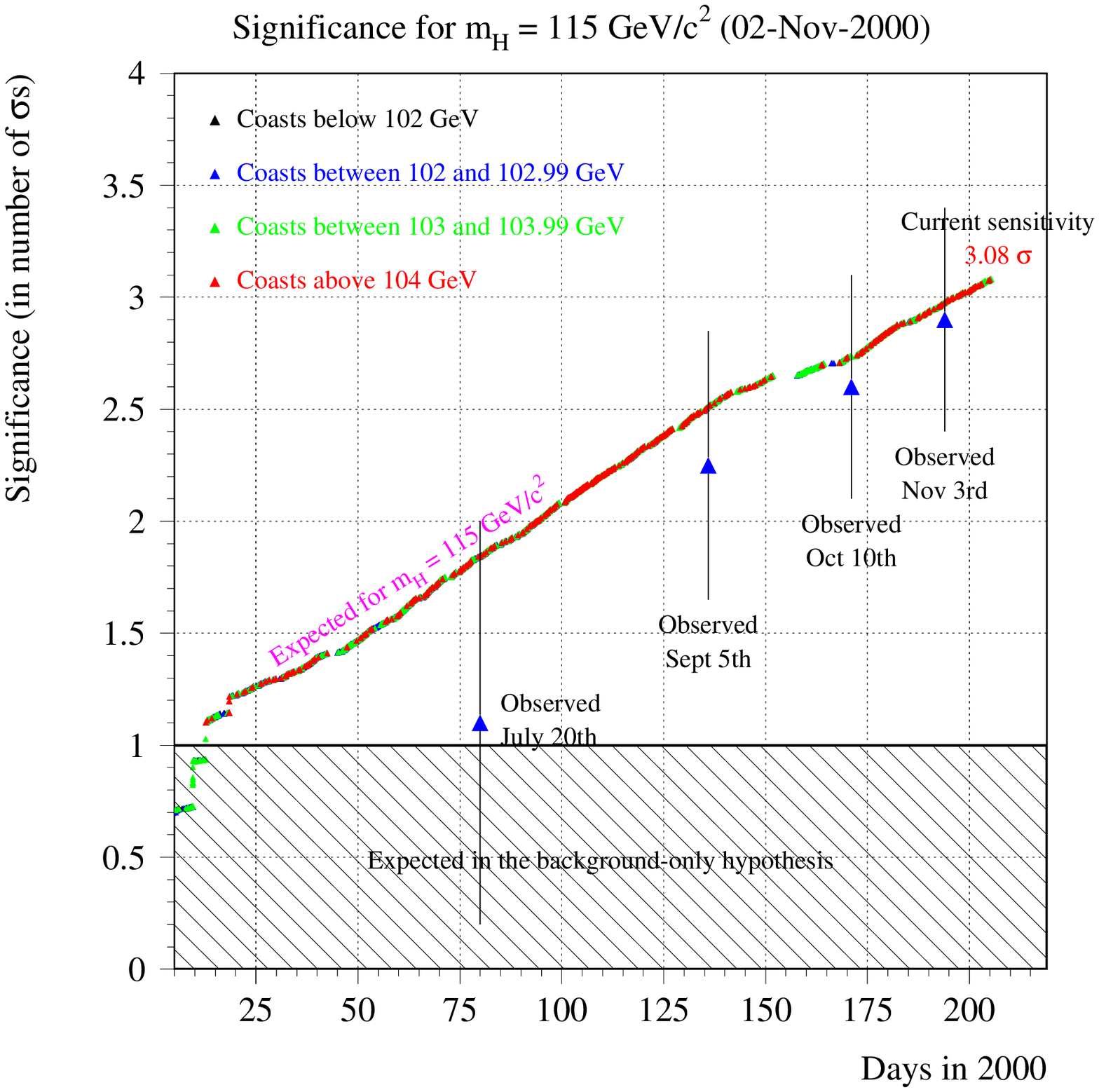}} 
\put(-1.8,5.8){(a)} &
\hskip.5cm\mbox{\epsfxsize=.45\hsize\epsfbox{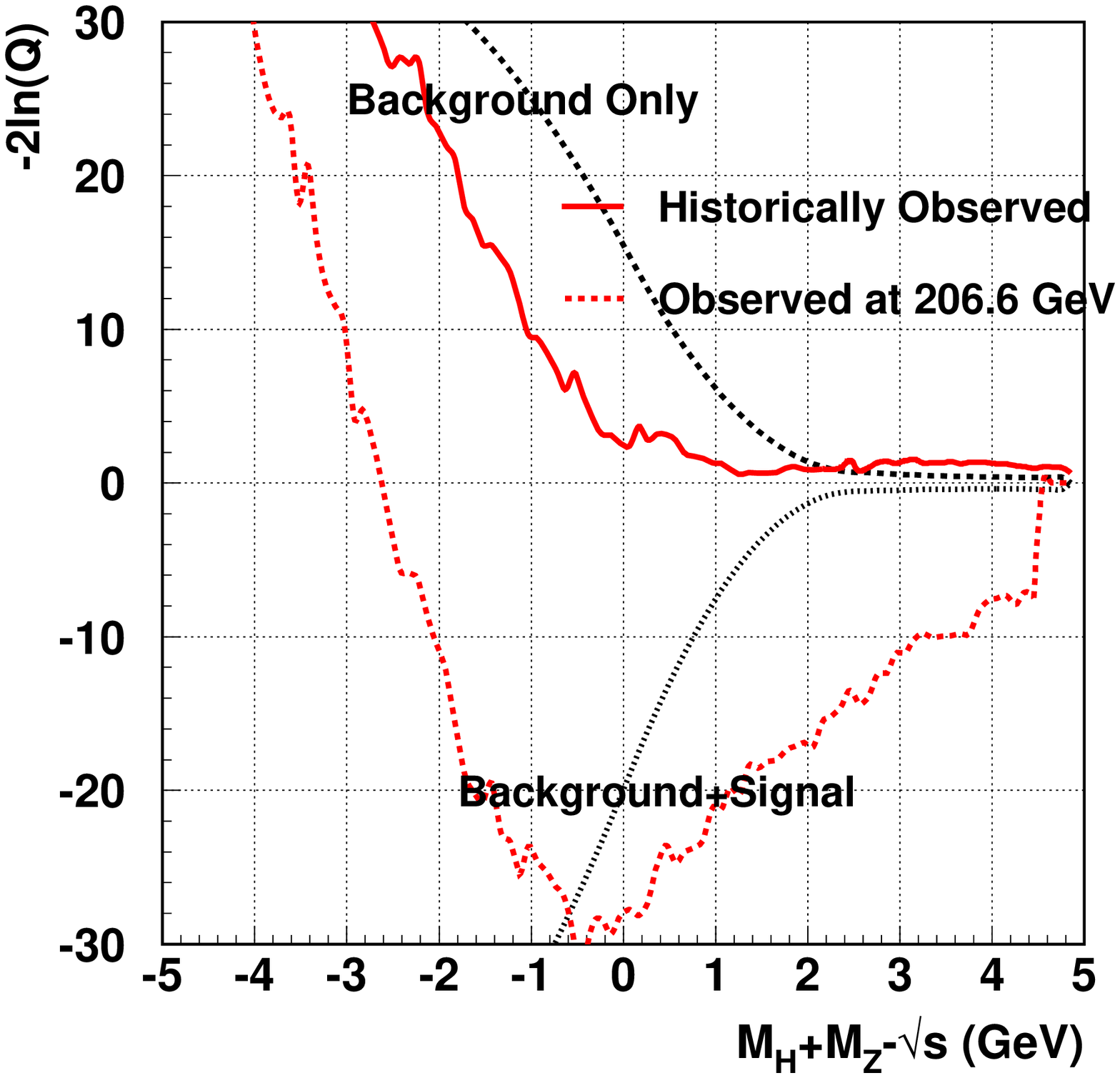}} 
\put(-1.8,5.8){(b)}
\end{tabular}
\vspace{-.2cm}
\caption{(a) Evolution of the observed significance at 
  $m_{\rm H}=115$~GeV/c$^2$ in 2000, compared to the expected increase
  in the signal hypothesis. The error bars are statistical only, with
  large point-to-point correlation. The expectation in the background
  only hypothesis is also indicated in the slanted region. (b)
  Observed estimators for the combination of all experiments and all
  data with $\sqrt{\rm s}<$206.5~GeV (full curve) and what it would
  have been had the excess observed above $\sqrt{\rm s}>$206.5~GeV
  been seen at all centre-of-mass energies (dashed curve) as a
  function of the distance of the Higgs boson mass hypothesis to the
  threshold. The expectation in absence and presence (dotted curves)
  of signal are also shown.
\label{llr}}
\end{center}
\end{figure}

To check the compatibility of the effect in different data samples
with various signal purities, the development of the excess is
compared to its expected evolution in the hypothesis of a
115~GeV/c$^2$ signal. As shown in Fig.~\ref{llr}-a, the constant
growth of the excess in time is consistent with the signal hypothesis.

To exclude the possibility of a systematic bias near the kinematic
threshold, a combination of the searches with the 500~pb$^{-1}$ of
data taken at centre-of-mass energies ranging from 189 to 206~GeV is
compared to what it would have been had the excess observed above
206~GeV been due to systematic bias close to the kinematic threshold.
The estimators, for both cases, are displayed in
Fig.~\ref{llr}-b~\cite{llrct} as a function of the distance to
threshold $(m_{\rm H}+m_{\rm Z}-\sqrt{s})$. This comparison shows how
dramatic the effect of a systematic bias would have been. Its absence
in the data taken at centre-of-mass energies below 206~GeV therefore
illustrates the robustness of the analyses near the kinematic
threshold.

\vspace{.25cm}
\noindent{\bf {\it Latest Updates}}
\vspace{.25cm}

In the remaining $\sim$30~pb$^{-1}$ of the last week of data taking,
only one significant event has been collected in OPAL in the four jet
channel. The significance for $m_{\rm H}~=~115$~GeV/c$^2$ is therefore
decreased by approximately 5\% in ALEPH and is unchanged in OPAL. The
significance of the DELPHI deficit is also unchanged. The L3
collaboration has not updated the significance of its excess.
Therefore, the conclusions with respect to those which can be drawn
from the combined results~\cite{lepc} are unchanged.

\vspace{.25cm}
\noindent{\bf {\it Conclusion}}
\vspace{.25cm}

In view of its consistency in all regards and its robustness, the
evidence for a 115~GeV/c$^2$ Higgs boson is as strong as could be
expected from the amount of data collected at centre-of-mass energies
above 206~GeV at LEP.

\vspace{.25cm}
\noindent{\bf {\it Acknowledgements}}
\vspace{.25cm}

It is a pleasure to thank the organisers and the secretariat of the
LEP3 conference in Rome for their wonderful hospitality. 
I am very grateful Jean-Baptiste de Vivie for his careful reading of these
proceedings.



\begin{thebibliography}{99}

\bibitem{lepc} P. Igo-Kemenes, {\it ``Status of Higgs boson
    searches''}, LEPC, november 2000.

\bibitem{request} LEP Higgs Working Group, {\it ``Standard Model Higgs
    Boson at LEP: Results with the 2000 data, Request for running in
    2001''}, http://alephwww.cern.ch/~janot/LEPCO/lephwg.ps

\bibitem{aleph}
ALEPH Collaboration,
{\it ``Observation of an excess in the search for the standard model Higgs  boson at ALEPH''},
Phys.\ Lett.\ B {\bf 495}, 1 (2000).

\bibitem{delphi} DELPHI Collaboration,
{\it ``Search for the standard model Higgs boson at LEP in the year 2000''},
Phys.\ Lett.\ B {\bf 499}, 23 (2001).

\bibitem{l3}
L3 Collaboration,
{\it ``Higgs candidates in e$^+$e$^-$ interactions at $\sqrt{\rm s}$=206.6~GeV''},
Phys.\ Lett.\ B {\bf 495}, 18 (2000).

\bibitem{opal} OPAL Collaboration,
{\it ``Search for the standard model Higgs boson in e$^+$e$^-$ collisions at  $\sqrt{\rm s}$=192 to 209~GeV''},
Phys.\ Lett.\ B {\bf 499}, 38 (2001).

\bibitem{llrct} LEP Higgs Working Group, {\it ``A Hint of the Standard Model Higgs Boson at LEP: Frequently Asked Questions and Their Answers''}, \\
  http://alephwww.cern.ch/~janot/LEPCO/qanda.ps

\end{thebibliography}
\end{document}